\documentclass[preprint2,12pt]{emulateapj}
\usepackage[]{graphicx}
\usepackage{amsmath}
\usepackage{amssymb}
\usepackage{mathrsfs}
\usepackage{subfigure}
\usepackage{boxedminipage}
\usepackage{aas_macros} 
\usepackage{natbib}
\usepackage{threeparttable}
\usepackage{appendix}
\usepackage{color,setspace,latexsym,amsthm,subfigure,rotating,longtable,arcs}
\usepackage{setspace}
\usepackage{Times}

\newcommand{\myr}{${\rm M_{\sun}\,yr^{-1}}$}
\newcommand{\Msol}{${\rm M_{\sun}}$}
\newcommand{\Lsol}{${\rm L_{\sun}}$}
\newcommand{\um}{$\mu$m}
\newcommand{\uJy}{$\mu$Jy}

\newcommand{\rxv}{$(r/r_{200})\times(\Delta v/\sigma_v)$}

\shorttitle{The Phase Space of $z\sim1.2$ Clusters with \textit{Herschel}}
\shortauthors{Noble et al.}

\begin{document}

\title{The Phase Space of $\lowercase{z}\sim1.2$ S\lowercase{p}ARCS Clusters: Using \textit{Herschel}${^\star}$ to probe Dust
Temperature as a Function of Environment and Accretion History}

\author{A.G. Noble\altaffilmark{1}, T.M.A. Webb\altaffilmark{2}, H.K.C. Yee\altaffilmark{1}, A. Muzzin\altaffilmark{3}, G. Wilson\altaffilmark{4}, R.F.J. van der Burg\altaffilmark{5}, M.L. Balogh\altaffilmark{6}, and D.L. Shupe\altaffilmark{7}}

\altaffiltext{$\star$}{{\it Herschel} is an ESA space observatory with science instruments provided by European-led Principal Investigator consortia and with important participation from NASA.}
\altaffiltext{1}{Department of Astronomy and Astrophysics, University of Toronto, 50 St George Street, Toronto, Ontario M5S 3H4, Canada}
\altaffiltext{2}{Department of Physics, McGill University, 3600 rue University, Montr\'{e}al, Qu\'{e}bec H3A 2T8, Canada}
\altaffiltext{3}{Institute of Astronomy, University of Cambridge, Madingley Rd, Cambridge CB3 0HA, UK}
\altaffiltext{4}{Department of Physics and Astronomy, University of California, Riverside, CA 92521, USA}
\altaffiltext{5}{Laboratoire AIM, IRFU/Service d'Astrophysique - CEA/DSM - CNRS - Universit Paris Diderot, Bat. 709, CEA-Saclay, 91191 Gif-sur-Yvette Cedex, France}
\altaffiltext{6}{Department of Physics and Astronomy, University of Waterloo, Waterloo, Ontario, N2L 3G1, Canada}
\altaffiltext{7}{NASA Herschel Science Center, IPAC, 770 South Wilson Avenue, Pasadena, CA 91125, USA}

\begin{abstract}
We present a five-band \textit{Herschel} study (100--500\,\um) of three galaxy clusters at $z\sim1.2$ from the Spitzer Adaptation of the Red-Sequence Cluster Survey (SpARCS).  With a sample of 120 spectroscopically-confirmed cluster members, we investigate the role of environment on galaxy properties utilizing the projected cluster phase space (line-of-sight velocity versus clustercentric radius), which probes the time-averaged galaxy density to which a galaxy has been exposed.  We divide cluster galaxies into phase-space bins of \rxv, tracing a sequence of accretion histories in phase space.  Stacking optically star-forming cluster members on the \textit{Herschel} maps, we measure average infrared star formation rates, and, for the first time in high-redshift galaxy clusters, dust temperatures for dynamically distinct galaxy populations---namely, recent infalls and those that were accreted onto the cluster at an earlier epoch.  Proceeding from the infalling to virialized (central) regions of phase space, we find a steady decrease in the specific star formation rate and increase in the stellar age of star-forming cluster galaxies.  We perform a probability analysis to investigate all acceptable infrared spectral energy distributions within the full parameter space and measure a $\sim4\sigma$ drop in the average dust temperature of cluster galaxies in an intermediate phase-space bin, compared to an otherwise flat trend with phase space.  We suggest one plausible quenching mechanism which may be consistent with these trends, invoking ram-pressure stripping of the warmer dust for galaxies within this intermediate accretion phase.

\end{abstract}


\section{Introduction}
\label{sec:intro}
In the framework of hierarchical structure formation, galaxy clusters continually build up their mass over time,  preferentially accreting matter along the cosmic filaments.  Thus, in the most basic picture, galaxy clusters consist of hundreds of galaxies belonging to one of two populations: an older collection of galaxies that may have formed in situ; and a younger population that was accreted over cosmic time.  This signifies the potential of clusters as laboratories with which to gauge differences between galaxies formed in distinct environments---the foundation of galaxy evolution studies.  However, it also necessitates an accurate definition for environment, as a cluster observed at a single redshift contains galaxies that have been exposed to different density environments dictated by their accretion time onto the cluster.

The significance of clusters as nurseries for galaxy transformations is substantiated by the many correlations between environment and galaxy properties, such as star formation rate (SFR), age, color and morphology \citep[e.g.,][]{Dressler97, Strateva01, Gomez03, Baldry04, Balogh04b, Kauffmann04, Hogg04, Blanton05, Poggianti08, Tran09, Finn10, Vulcani10, Muzzin12}.  However, there also exists a strong covariance between stellar mass and environment, as massive galaxies are found in increasingly denser regions \citep{Kauffmann04, Baldry06}.  The key to disentangling this covariance requires systematically mapping out trends with each parameter, while fixing the other, over cosmic time. Indeed, once the star-formation history is accounted for, either through stellar age or mass, many of the aforementioned environmental trends appear to weaken \citep[e.g.,][]{Blanton09,Peng10, Muzzin12, Wetzel12}.  

Further elucidation of the processes that govern galaxy evolution relies primarily on two criteria: the extension of cluster surveys beyond $z\sim1$, and a coherent definition of environment.  Indeed, it is now clear that the peak epoch of star formation occurred at $1<z<3$ \citep[e.g.,][]{Madau96, Hopkins04, Hopkins06, Bouwens07, Madau14}.  Moreover, the fraction of star-forming galaxies within clusters increases with redshift, as seen in optical \citep{BO78, Ellingson01} and infrared studies \citep{Saintonge08, Haines09, Finn10, Webb13, Alberts14} out to $z\sim1$, and may even rise with increasing galaxy density in clusters beyond $z\sim1.5$ \citep{Tran10, Brodwin13, Ma15, Santos15}.  This evolution in star-forming galaxies appears to be dominated by the infalling population as it mimics the changes in the coeval field population \citep{Haines09, Webb13}.  As such, a proper definition of environment that isolates the recently accreted infalling population from the older in-situ population is crucial to accurately assess the effect of environmental quenching.  

A galaxy's path taken through a cluster, and thus its exposure to different density environments, is encoded in its orbital history.  Unfortunately, this is not directly observable as we are limited to a single projected snapshot in time.  Recent simulations, however, have shown that cluster phase space---member galaxies' line-of-sight velocity relative to the cluster versus clustercentric radius---can help to circumvent this problem as it is sensitive to the time since galaxy infall \citep{Gill05, Haines12, Taranu14}.  Moreover, distinct regions in phase space can isolate different satellite populations \citep{Mahajan11, Oman13}.

Utilizing the Millennium Simulation from \cite{Springel05}, \cite{Haines12} trace out galaxy accretion histories for orbiting galaxies of 30 massive clusters as a function of phase space (see figure~3 in \citealp{Haines12}).  These diagrams reveal the distinct trumpet- (or chevron-) shaped loci occupied by the older, virialized population with respect to the more recently accreted infall population (see also \citealp{Haines15}).  This latter population consists of galaxies that have not yet entered into the virial cluster radius, those that have already reached pericenter and are on their way back out, known as back-splash galaxies \citep{Balogh00, Mamon04}, and everything in between.  Thus, a phase-space analysis for environment can effectively account for distinctive cluster populations and alleviate some of the projection effects that bias the traditional probes for environment: clustercentric radius and local density. 

Many recent studies have exploited phase space to further study galaxy properties, such as morphologies \citep{Biviano02}, AGN distribution \citep{Haines12}, post-starburst distribution \citep{Muzzin14}, optical colors \citep{Crawford14}, star formation activity \citep{Hernandez14}, HI gas \citep{Jaffe15}, and quenching timescales \citep{Haines15}.  In \cite{Noble13_240}, we parameterized the accretion history of a $z=0.871$ cluster with lines of constant \rxv\ in phase space to study environmental effects on 24\um-detected cluster galaxies.  These lines roughly delineate the trumpet-shaped caustic regions that trace the expected orbital velocities within massive clusters \citep{Regos89}.  Once we group galaxies according to their accretion history in \cite{Noble13_240}, we find an environmental dependence on stellar age and specific star formation rate (SSFR) for star-forming galaxies, rather than the flatter trend seen with clustercentric radius and galaxy density \citep[e.g.,][]{Blanton09,Peng10, Muzzin12, Wetzel12}.  The parameterization of \rxv\ thus has the potential to expose environmental processes that could otherwise be hidden due to the mixing of infall histories.

In this paper, we extend the phase-space analysis to three spectroscopically-confirmed clusters at $z\sim1.2$.  The improvements to \cite{Noble13_240} are primarily two-fold: we consider a more statistically significant sample of 123 cluster members (48 of which are optically star-forming) and we measure more robust SFRs through full coverage of the thermal portion of the spectral energy distribution with five-band \textit{Herschel} photometry.

In Section~\ref{sec:reduc} we describe the sample and the \textit{Herschel} observations.  We discuss our stacking analysis in Section~\ref{sec:analysis}, and the results from the phase-space analysis in Section~\ref{sec:results}.  We propose a plausible quenching scenario in Section~\ref{sec:disc}, with our final remarks in Section~\ref{sec:conclusions}.   We assume a standard cosmology throughout the paper with $H_0=70$\,km\,s$^{-1}$\,Mpc$^{-1}$, $\Omega_{\textup{M}}=0.3$, $\Omega_{\Lambda}=0.7$.  Stellar masses and SFRs are based on a Chabrier initial mass function \citep{Chabrier03}.

\section{Observations and Data Reduction}
\label{sec:reduc}

\subsection{A $z\sim1.2$ Sample from SpARCS/GCLASS}
\label{sec:gclass}
The three clusters in this study derive from the Spitzer Adaptation of the Red-Sequence Cluster Survey (SpARCS) \citep{Wilson09, Muzzin09, Demarco10}, which utilizes an infrared color technique to pinpoint the red sequence of cluster galaxies \citep{Gladders00} out to $z\sim1.6$.  Specifically, SpARCS locates over-densities of red-sequence cluster galaxies using a $z-3.6$\um\ color selection which brackets the 4000\,\AA-break at $z>1$.  This simultaneously traces dense cluster regions while providing a photometric redshift for the cluster from the color of the red-sequence.  Additionally, the three $z\sim1.2$ clusters were selected as part of an ambitious spectroscopic follow-up survey, the Gemini Cluster Astrophysics Spectroscopic Survey (GCLASS; \citealp{Muzzin12}).  Utilizing the Gemini Multi-Object Spectrograph (GMOS), GCLASS has successfully obtained redshifts for $\sim$800 galaxies within ten rich cluster fields from $0.85<z<1.3$, including $\sim400$ cluster members.

To optimize the number of cluster redshifts, GCLASS targets were prioritized based primarily on three criteria: small clustercentric radius; proximity in observed $z^\prime - 3.6$\um\ color to the cluster red-sequence; and 3.6\um\ flux.  A 3.6\um\ flux-selected sample is advantageous at $z\sim1$ as it probes rest-frame $H$ band and is thus similar to a stellar-mass limited sample. The color selection was broader at $z\sim1.2$ to account for the bluer rest-frame probed by the red-sequence color.  This helps to eliminate any potential selection biases by spanning the colors of both red and blue cluster members, while excluding obvious background/foreground galaxies. Additional priority was also given to galaxies off the red-sequence with 24\um-MIPS detections to select dusty star-forming cluster member candidates without any specific color cut.   Cluster members are defined as sources within 1500 km\,s$^{-1}$ of the cluster velocity dispersion, yielding a final sample of 122 cluster members with secure redshifts within the three $z\sim1.2$ clusters presented here.  Cluster properties, including M$_{200}$, R$_{200}$, and $\sigma_v$ (G. Wilson et al., in preparation) are listed in Table~\ref{tab:z1_clusts}, and spectroscopic details, target selection, survey completeness, and stellar masses are described in detail in \cite{Muzzin12}.

\begin{table*}
\begin{center}
\caption{The \textit{Herschel}-GCLASS sample at $z\sim1.2$.}
\label{tab:z1_clusts}
\begin{tabular}{ccccccccc}
\hline

\multicolumn{1}{c}{SpARCS Name} &
\multicolumn{1}{c}{Nickname} &
\multicolumn{1}{c}{$z_{spec}$} &
\multicolumn{1}{c}{RA} &
\multicolumn{1}{c}{Dec} &
\multicolumn{1}{c}{M$_{200}$} &
\multicolumn{1}{c}{R$_{200}$} &
\multicolumn{1}{c}{$\sigma_v$} &
\multicolumn{1}{c}{\# of} \\
\multicolumn{1}{c}{} &
\multicolumn{1}{c}{} &
\multicolumn{1}{c}{} &
\multicolumn{1}{c}{(J2000)} &
\multicolumn{1}{c}{(J2000)} &
\multicolumn{1}{c}{(10$^{14}$\Msol)} &
\multicolumn{1}{c}{(Mpc)} &
\multicolumn{1}{c}{(km s$^{-1}$)} &
\multicolumn{1}{c}{CMs} \\

\hline
J161641+554513 & EN1-349 & 1.1555 & 16 16 41.232 & $+$55 45 25.708 & 1.7$^{+0.7}_{-0.7}$ & 0.74$^{+0.09}_{-0.12}$ & 660$^{+80}_{-110}$ & 42\\
J163435+402151 & EN2-111 & 1.1771 & 16 34 35.402 & $+$40 21 51.588 & 3.1$^{+1.3}_{-1.1}$ & 0.89$^{+0.11}_{-0.12}$ &  810$^{+100}_{-110}$ & 43\\
J163852+403843 & EN2-119 & 1.1958 & 16 38 51.625 & $+$40 38 42.893 & 0.77$^{+0.31}_{-0.40}$ & 0.56$^{+0.06}_{-0.12}$ & 510$^{+60}_{-110}$ & 37\\
\hline

\end{tabular}
\end{center}
\end{table*}

\subsection{\textit{Herschel}-PACS Imaging}
\label{sec:pacs}
We observed with the Photodetector Array Camera and Spectrometer (PACS) instrument \citep{Poglitsch10} aboard the \textit{Herschel} Space Observatory \citep{Pilbratt10} over 5$\times$5 arcmin around each cluster using the medium speed scan (20 arcsec/s) in array mode with homogenous coverage.  We divided each cluster into two astronomical observation requests (AORs) with alternating scan directions of $45^\circ$ and $135^\circ$ to further maximize homogeneity (OBSIDs 1342247340, 1342247341 for EN1-349; 1342248661, 1342248662 for EN2-111; and 1342248631, 1342248632 for EN2-119).  The final maps consist of 7.2 hours of integration over each field at 100 and 160 \um.

\begin{figure*}[] \centering

\subfigure{\includegraphics[width=7.5cm]{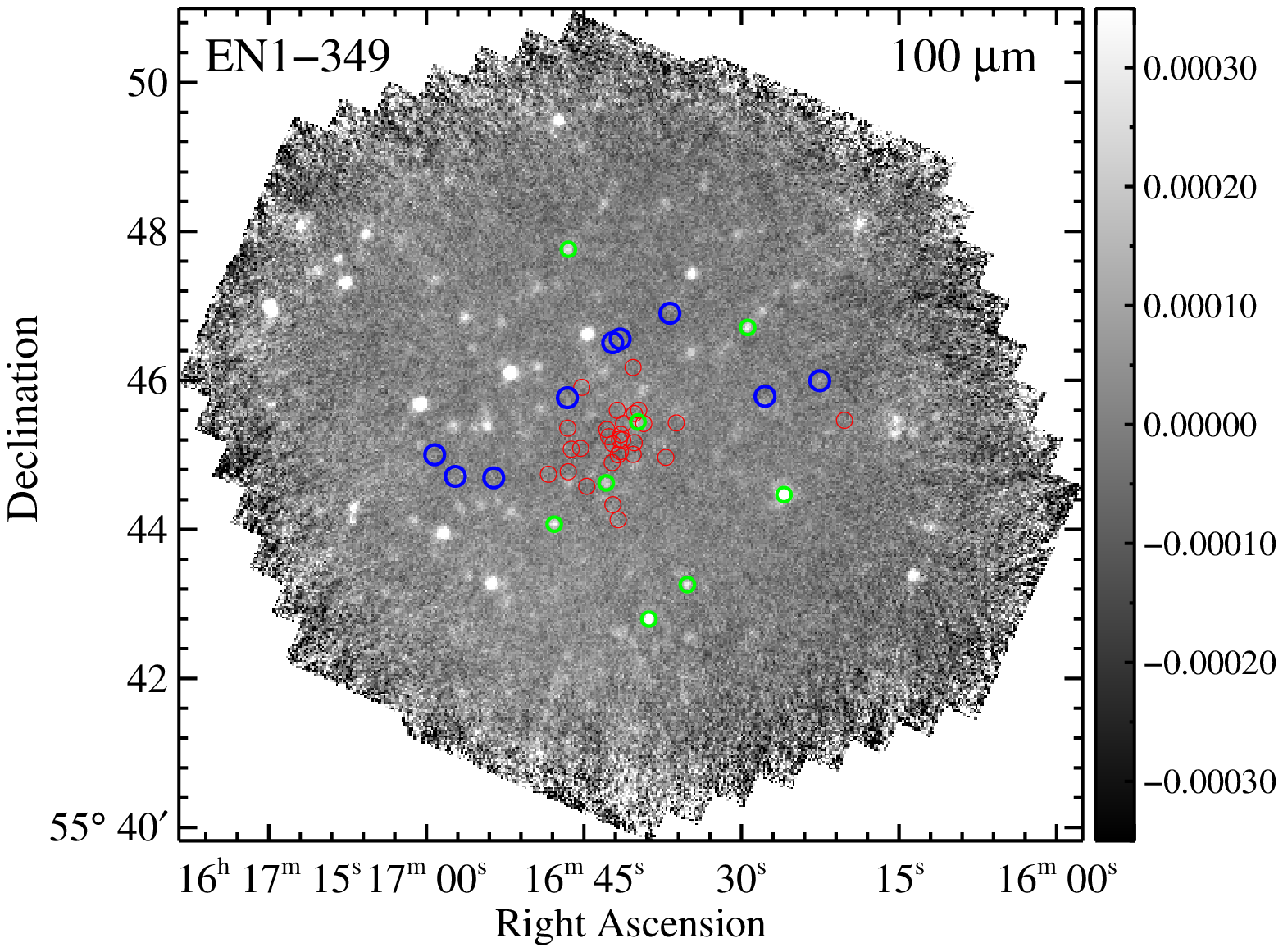}}
\subfigure{\includegraphics[width=7.5cm]{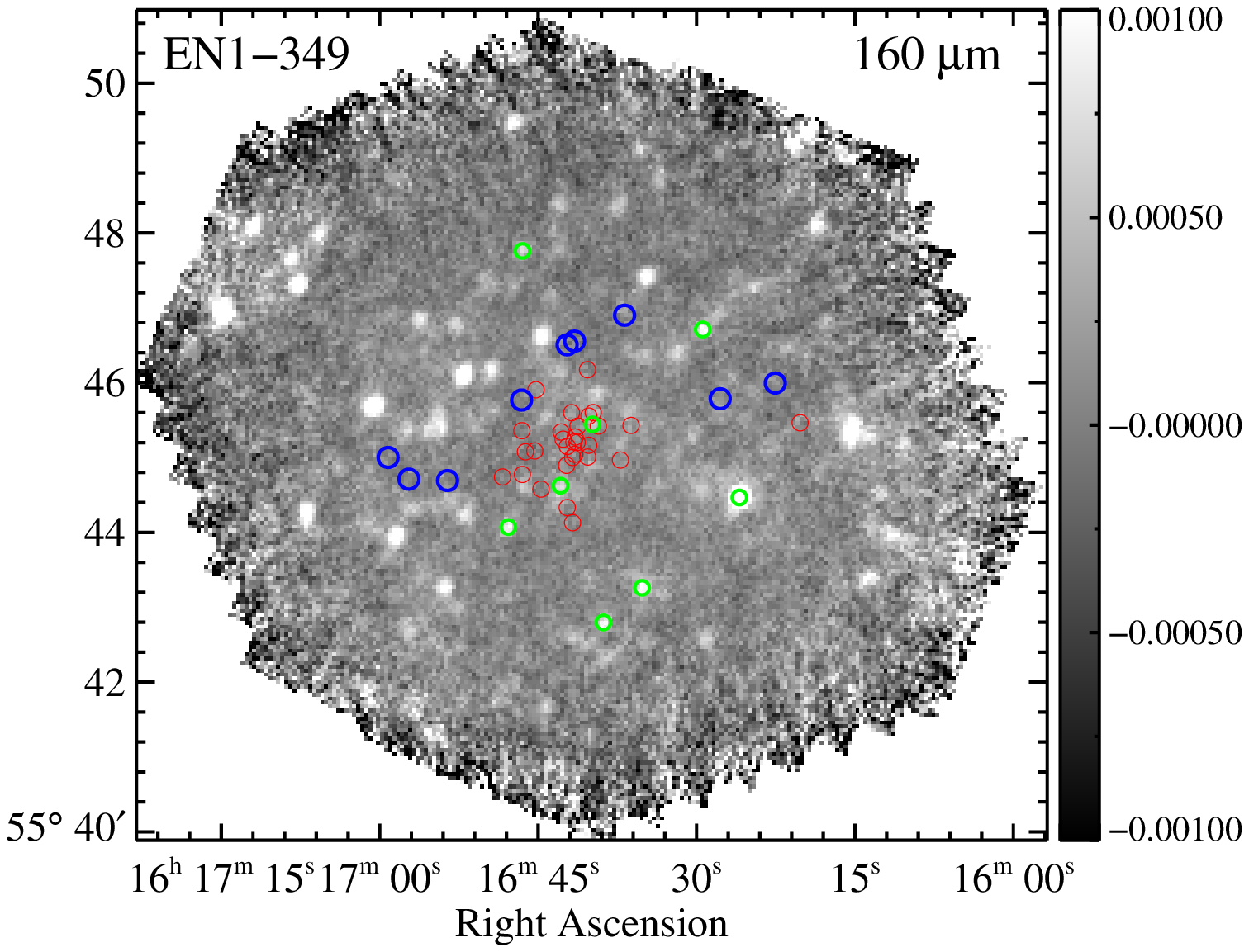}}

\vspace{-4.mm}

\subfigure{\includegraphics[width=7.5cm]{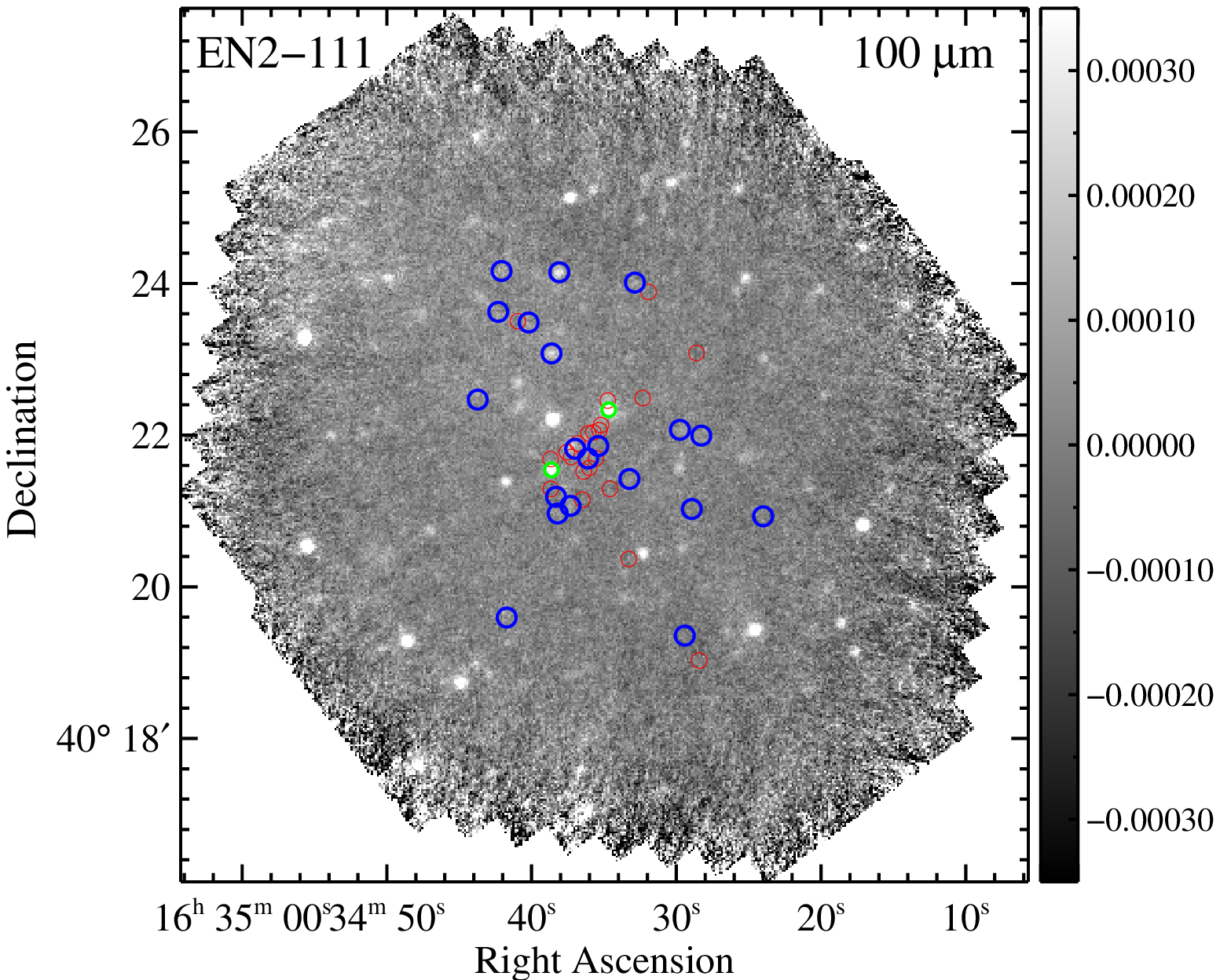}}
\subfigure{\includegraphics[width=7.5cm]{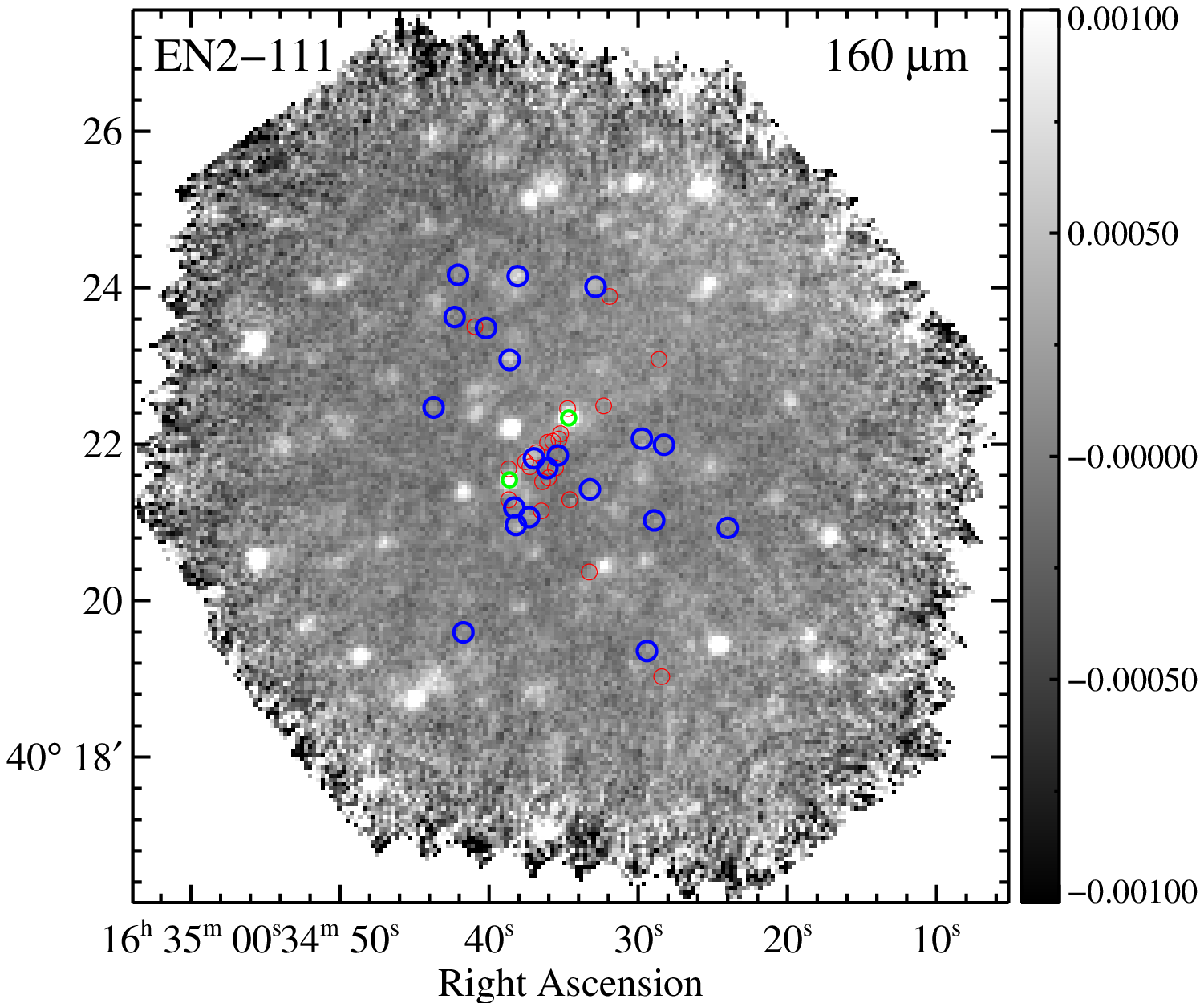}}

\vspace{-2.mm}

\subfigure{\includegraphics[width=7.5cm]{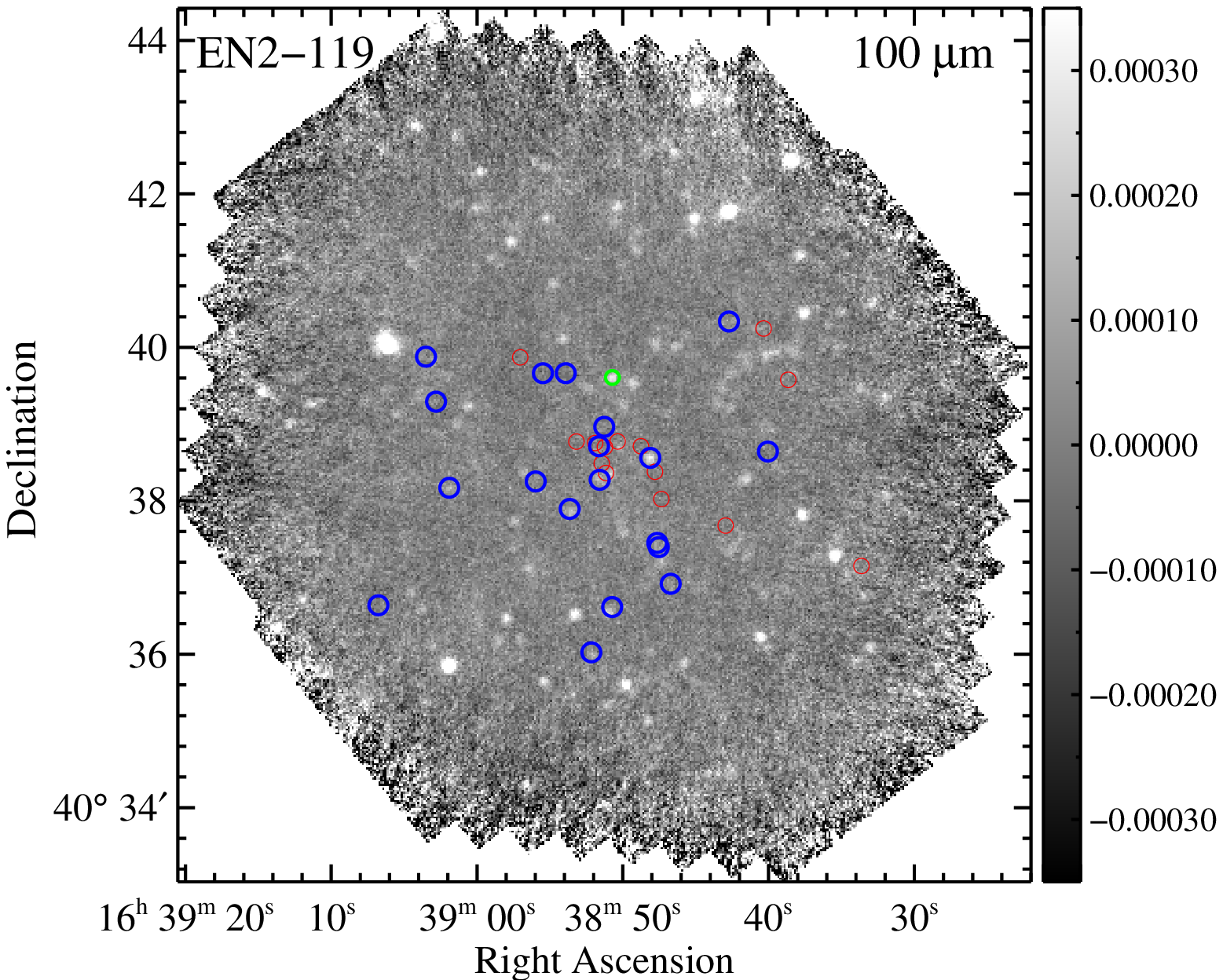}}
\subfigure{\includegraphics[width=7.5cm]{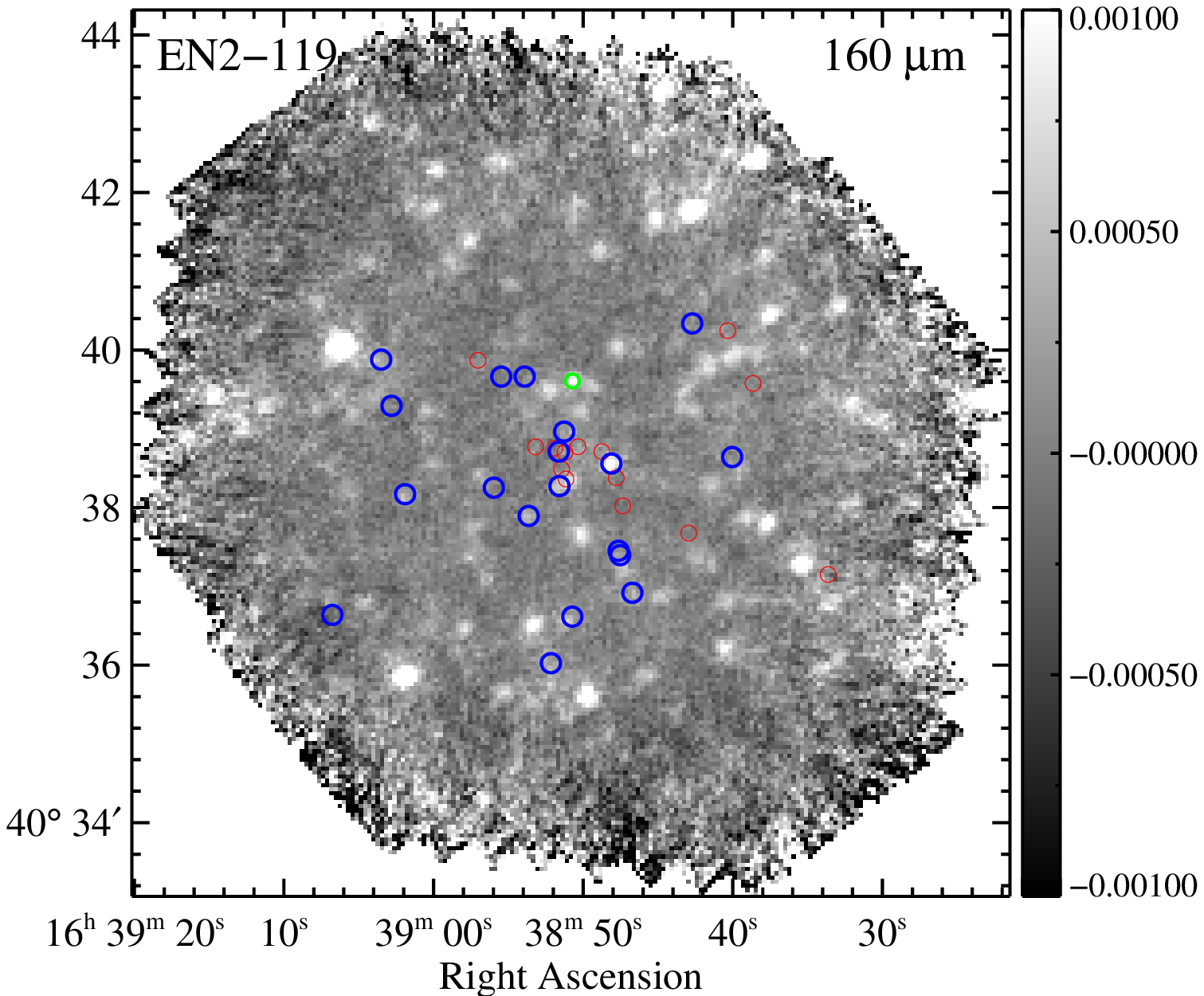}}

\vspace{-0mm}
\caption{\textit{Herschel}-PACS maps at 100 (left column) and 160 \um\ (right column) for EN1-349 (top row), EN2-111 (middle row) and EN2-119 (bottom row).  We plot spectroscopically confirmed cluster members with [OII] emission in blue and cluster members without [OII] emission in red.  In green, we additionally highlight foreground and background galaxies (i.e., non-cluster galaxies) that have high-quality spectroscopic redshifts and are individually detected in the PACS maps.}\label{fig:pacs_maps} 
\end{figure*}

We reduce each map with the Unimap pipeline \citep{Piazzo15}, which employs a generalized least squares map-making technique to help reduce the ubiquitous $1/f$ noise.  We first produce Level 1 products from the raw AORs using the automatic pipeline in the {\sc hipe v12.0} environment \citep{Ott10}.  We then use UniHipe to convert these products into fits files appropriate to input into Unimap.  We run Unimap with the standard parameters and project onto final pixel sizes of 1.6 and 3.2\arcsec\ for 100 and 160\um, respectively.  In Figure~\ref{fig:pacs_maps} we show the maps for each cluster field in both channels, over-plotted with symbols denoting the positions of spectroscopically confirmed star-forming cluster members.

\subsection{\textit{Herschel}-SPIRE Data}
\label{sec:spire}
We download raw archival data from the Herschel Science Archive for the Spectral and Photometric Imaging Receiver (SPIRE; \citealp{Griffin10}) for the Elais North-1 and Elais North-2 fields, which contain our three clusters, and reduce the data ourselves.  The observations derive from the Herschel Multi-tiered Extragalactic Survey (HerMES), the largest program carried out with \textit{Herschel}, covering $\sim$70 deg$^2$ \citep{Oliver12}.  Both fields belong to the level-6 tier of the survey, meaning they are wide-field but relatively shallow depth, and were observed in parallel mode with PACS and SPIRE.  The relevant obsids are 1342228450/1342228354 for Elais North-1 and 1342214712/1342226997 for Elais North-2. 

In general, we follow the default pipeline for reducing large SPIRE maps in parallel mode, with a few modifications when necessary.   Processing each AOR separately, the pipeline first loops over the scans, correcting for various artifacts, including jumps in the thermistor timelines, cosmic ray glitches, low-temperature noise drifts, and pointing calibrations. We then merge the orthogonal scan directions from each AOR.

The next step involves making a correction for residual offsets between detectors, which can lead to stripes in the map.  The SPIRE Destriper module has been found to produce optimal results \citep{Xu14}.  This algorithm iteratively removes the offsets by adjusting the baseline removal with either a simple median or polynomial fit of a specified degree.  We use a zeroth-order polynomial for the Elais North-2 field, and a first-order polynomial for the Elais North-1 field, which was found to have significant stripes with the zeroth-order fit.  While a higher-order polynomial has been found to produce a tilted background in some cases, visual inspection of the resulting map revealed it was a robust solution, producing no varying extended emission.  Finally, the baseline-removed scans are passed to the map maker; again we use the default naive mapper, which projects each bolometer signal onto a sky pixel, and creates a flux density map by dividing the total signal map by the coverage map.  

\section{Analysis}
\label{sec:analysis}
Since most of the cluster galaxies are undetected in the PACS and SPIRE maps, we perform a stacking analysis (see Sections~\ref{sec:pstack} and \ref{sec:sstack}) to study the average properties of cluster members.  We utilize the extensive spectroscopy available over all three clusters (122 members), but focus on non-passive galaxies (defined by the presence of [OII] emission), as we are primarily interested in environmental effects on star formation activity.  This is a reasonable selection given that GCLASS spectroscopy in $z\sim1$ galaxy clusters is not significantly biased against dusty star formation, as seen through an agreement of the total SFR per unit cluster mass measured separately by the [OII] equivalent width and 24\,\um\ flux \citep{Webb13}.  Moreover, we note that 78\% of 24\um-detected cluster members have [OII] emission over the three $z\sim1.2$ clusters.

After removing quiescent galaxies from the sample, we are left with 57 star-forming galaxies.  Through visual inspection, we further remove 9 sources that are heavily contaminated by a bright (at 100 or 160\um), nearby field galaxy, yielding 9, 20, and 19 star-forming cluster galaxies in EN1-349, EN2-111, and EN2-119, respectively.  We further divide the sources into four phase-space bins, defined as \rxv\ as described in \cite{Noble13_240}, with 12 galaxies per bin. 

\subsection{PACS Stacking}
\label{sec:pstack}
For each bin, we stack the PACS maps at 100 and 160 \um\, combined over all three clusters, using the following approach.  We extract a thumbnail image around each source, large enough to cover the recommended annulus for sky estimation (see Section~\ref{sec:pacs_phot}), yielding a radius of 26 arcsec for 100 \um\ and 30 arcsec for 160 \um.  We mask the inner pixels of the thumbnails (6 and 12 arcsec), just slightly more than the recommended radius for aperture photometry at each wavelength, and perform a $3\sigma$ clipping on the remaining pixels to remove any bright sources within the extracted map, thereby cleaning the sky annulus region.  We also remove any $2\sigma$ outliers that neighbor the $3\sigma$ pixels. The clipped map is then weighted by the inverse of the variance taken from a matched thumbnail of the noise map (the standard deviation of the naive map).  We combine all subsequent thumbnails to make a cube for each bin, where each layer represents the clipped, weighted map with the inner masking now removed.  Along the cube dimension, we perform a trimmed average: we first discard the minimum and maximum pixel values along all layers in the cube (not including the clipped values), pixel by pixel, and calculate the mean in the remaining pixels.  Finally, we normalize by the total of the inverse variance associated with the pixels that contribute to the trimmed mean.  The resulting map, a flattened cube that has been averaged (with trimming) pixel by pixel with each layer inverse weighted and $3\sigma$ clipped, represents our stacked image.

\subsection{PACS Aperture Photometry}
\label{sec:pacs_phot}
We perform aperture photometry on the final stacked images within the {\sc hipe v12.0} environment, using the recommended radii for faint sources: 5.6, 20, and 25 arcsec (10.5, 24, and 28 arcsec) for source flux, inner sky annulus, and sky annulus, respectively at 100 \um\ (160 \um).  The sky estimate is determined from an algorithm adapted from {\sc daophot} which iteratively computes the mean and standard deviation of the provided sky pixels, each time removing possible outliers, and is ultimately subtracted from the source flux.

The aperture radii are below the nominal FWHM of the beam at each wavelength, which helps to reduce contamination from any residual flux from neighboring sources that was not removed during stacking.  We must then apply an aperture correction to properly account for the missing flux, calculated with version 7 of the responsivity function in each band: 0.57 for 5.6 arcsec at 100 \um\ and 0.64 for 10.5 arcsec at 160 \um. We bootstrap the sources 1000 times in each phase-space bin and calculate the standard deviation on the mean to determine the errors on the binned fluxes.

\subsection{SPIRE Stacking}
\label{sec:sstack}

The large beam at SPIRE wavelengths (18.2, 24.9, and 36.3 arcsec FWHM at  250, 350, and 500 \um, respectively) renders traditional stacking analyses difficult.  This is due to the confusion-limited nature of the maps, which is typically reached when the source density surpasses $\sim$0.02--0.03 beam$^{-1}$ \citep{Condon74}; this is found to occur at 18.7, 18.4, and 13.2 mJy (at 40 beams per source) with ascending SPIRE wavelengths \citep{Oliver10}.  As a result, a single flux peak in the map is likely to have contributions from a sea of fainter, unresolved sources and disentangling these fluxes can be problematic.  Moreover, this effect amplifies when one considers intrinsically correlated populations, as they are expected to be clustered on large beam scales.

Monte Carlo simulations can circumvent this problem by correcting for the bias measured in stacked fluxes.  However, an even cleaner approach lies in fitting the stacked fluxes of correlated populations simultaneously.  An algorithm called \textsc{simstack} developed specifically for SPIRE maps by \cite{Viero13} exploits the latter technique and has been made publicly available.  It was designed specifically to deconvolve the flux from inherently clustered populations (e.g., the CIB), and simulations found it to be an unbiased estimator compared to traditional stacking methods.

The general method relies on using positional priors from a deeper, less-confused image (e.g., 24\,\um), separating potentially clustered populations into individual lists, and fitting the flux at all positions simultaneously.  This effectively allows for a deconvolution of the flux contribution from multiple sources within a beam (assuming that all these sources were detected in the prior catalog). 

More specifically, a ``hits map" is created for each list of grouped sources, where pixels are assigned an integer value corresponding to the number of sources that fall within it.  These maps are smoothed with the FWHM of the beam and mean-subtracted.  For each list, a vector is populated with the values in the mean-subtracted smoothed map at all pixels of interest around each source, combined from all lists.  All vectors are then passed to the fitting routine of the functional form 
\begin{equation}
M_j = \sum_{\alpha}S_{\alpha}C_{\alpha j}, 
\end{equation}
where $M_j$ corresponds to the real map values in $j$ pixels of interest, $C_{\alpha j}$ is the beam-convolved mean-subtracted values in the pixels of interest for each list $\alpha$, and $S_{\alpha}$ is the stacked flux in list $\alpha$.  This process is iterated until the resulting $\chi^2$ is minimized, yielding a simultaneous stacked flux for each list.  The entire procedure is subsequently repeated for each SPIRE wavelength. 

As we aim to combine sources from all three clusters into each phase-space bin, we slightly alter the public code to suit our needs.  Specifically, we create a separate beam-convolved mean-subtracted ``hits map" for each field, and then merge the resulting vectors together (along with merged vectors of real data) to pass to the fitting routine.

In order to optimize the functionality of \textsc{simstack} for reducing the bias due to beam-size clustering, we simultaneously pass prior catalogs to the fitting routine that could contribute any source confusion (in addition to our spectroscopic catalogs).   MIPS-24\,\um\ \citep{Rieke04} is the conventional prior catalog to use for SPIRE for two reasons: it correlates well with far-infrared wavelengths, and a large fraction of 24\,\um\ sources are resolved \citep{Papovich04}.  We use deep MIPS maps for each field from a Guaranteed Time Observer program (proposal ID 50161) with exposures of 1200 seconds per pixel.  The catalogs are complete down to $\sim70$\,\uJy.  Our final simultaneously-stacked bins include: (1) four phase-space bins with star-forming ([OII]-detected) cluster members; (2) a single bin containing all remaining cluster members without [OII] emission; (3) two separate catalogs of field galaxies within the redshift range $1.10<z<1.21$, one containing [OII]-detected galaxies, and one with passive galaxies; and (4) all  24\,\um-detected sources above $3\sigma$.  This latter list has been purged of any cluster members and field galaxies that have already been included in the previous bins.  Error bars are computed from the standard deviation on 1000 bootstrapped means for each bin, stacking all catalogs listed above simultaneously.

\subsection{Possible AGN Contamination}
Recent studies have shown that high-redshift clusters harbor a higher fraction of AGN than their local counterparts \citep{Eastman07, Galametz09, Martini09}, and can reach up to 10\% at $z\sim1.25$ \citep{Martini13}.   AGN heat their surrounding dusty torus, which radiates monotonically in the mid-infrared with a power-law spectrum.  Thus, \textit{Spitzer}-IRAC colors can purge AGN from star-forming galaxies which display the 1.6\um\ stellar bump \citep{Lacy04, Stern05}. We exploit this technique to identify any cluster AGN in our sample using the color criteria from \cite{Lacy07}.  The IRAC photometry for GCLASS is described in \cite{vanderburg13}.  Only one cluster member with [OII]-emission falls into the AGN wedge, and is still consistent with the star-forming region within 1$\sigma$.  It also lies well outside the revised bounds proposed by \cite{Donley12} and \cite{Kirkpatrick13}, which further remove high-redshift star-forming interlopers.  We therefore choose to keep the source in the sample, as its infrared luminosity is unlikely to be dominated solely by an AGN.  

\section{Phase-Space Results}
\label{sec:results}
\begin{figure}[h] \centering
\includegraphics[width=8.5cm]{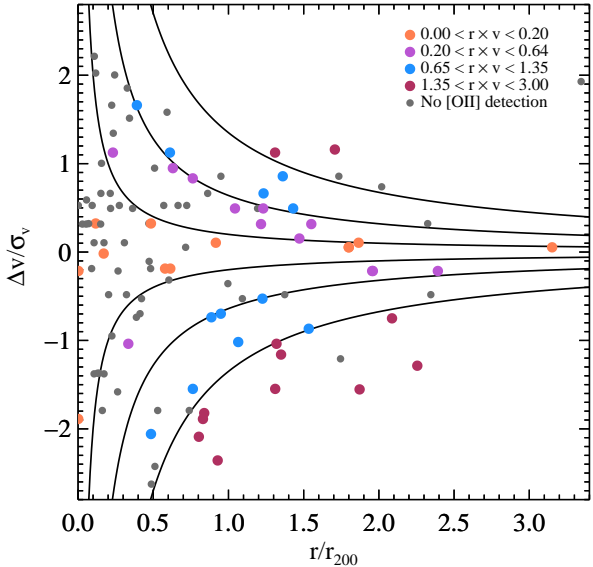}
\caption{The phase space (line-of-sight velocity versus clustercentric radius, normalized by the cluster velocity dispersion and r$_{200}$, respectively) for all cluster galaxies at $z\sim1.2$.  The gray points are galaxies without any discernible [OII] emission, while the larger colored points all have [OII] emission.  They are color-coded based on their position in phase space, with delineations of constant \rxv\ shown by the black lines. } 
\label{fig:phase} 
\end{figure}

In total, we utilize 48 spectroscopically-confirmed, star-forming cluster galaxies with [OII] emission.  Here, we extend the phase-space analysis from \cite{Noble13_240} to higher-redshifts and utilize more extensive photometry. Since we have stacked the Herschel fluxes, we aim to create phase-space bins with equal numbers of galaxies and therefore deviate slightly from the \rxv\ values used in \cite{Noble13_240}.  Our final bin delineations are as follows: $(r/r_{200})\times(\Delta v/\sigma_v) <0.20$ (central bin); $0.20<(r/r_{200})\times(\Delta v/\sigma_v) <0.64$ (intermediate bin); $0.64<(r/r_{200})\times(\Delta v/\sigma_v) <1.35$ (recently accreted bin); and $(r/r_{200})\times(\Delta v/\sigma_v) >1.35$ (infalling bin).  The resulting bins are shown in Figure~\ref{fig:phase}.

\subsection{Stellar Age and Mass in Phase Space}
\label{sec:phase}
\begin{figure}[h] \centering
\includegraphics[width=8.5cm]{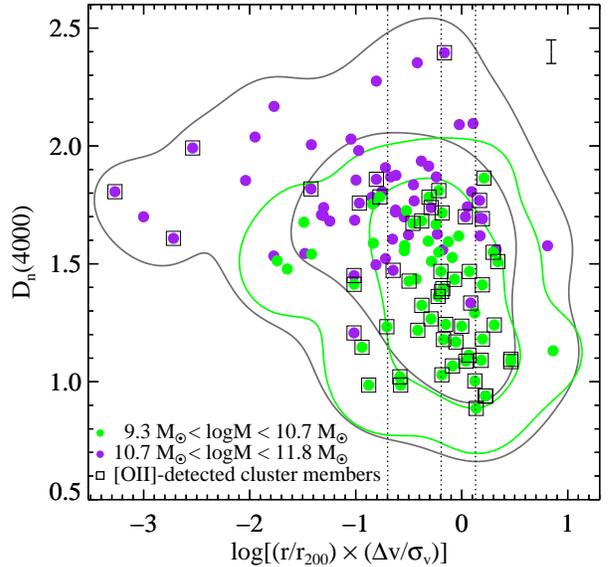}
\caption{The strength of the 4000\,\AA\-break as a function of $\log[(r/r_{200})\times(\Delta v/\sigma_v)]$ for all cluster members.  The green and purple circles represent galaxies with lower and higher stellar masses, respectively.  The open squares denote star-forming galaxies detected with [OII] emission.  The dashed vertical lines show the bin delineations we use in \rxv. The gray contours represent the 68\% and 95\% smoothed surface density regions after applying a two-dimensional kernel density estimator to all the points, while the green contours represent the lower mass galaxies only. A typical error bar is shown in the upper right corner.} 
\label{fig:d4000} 
\end{figure}

We test the efficacy of using phase space as an accretion history sequence by plotting the strength of the 4000\,\AA-break as a function of $\log[(r/r_{200})\times(\Delta v/\sigma_v)]$ for all cluster members in Figure~\ref{fig:d4000} (star-forming cluster members are highlighted with a box).  There is a general trend toward lower 4000\,\AA-break depths for higher values of \rxv.  A Spearman's $\rho$ test reveals a mild correlation of $-0.38$, with a high probability ($>99\%$) of rejecting the null hypothesis of no correlation.  This signifies that galaxies that were accreted most recently to the cluster have had a more recent episode of star formation.  After dividing the cluster galaxies into two stellar mass bins, defined by the median mass, the trend persists only for the lower mass bin with a mild correlation coefficient of $-0.32$ at high significance ($99\%$).  The null hypothesis is not rejected for the higher mass galaxies, suggesting that the trend is primarily driven by the low-mass galaxies.  While the two mass bins themselves are wide, covering over an order of magnitude of stellar mass,  the median mass in each bin remains roughly constant as a function of phase space, changing by less than 0.4 dex over each bin. This indicates that the relationship between the 4000\,\AA-break and phase space is not driven entirely by mass segregation, and the cluster environment may play a role at fixed (lower) stellar mass.  We note that 74\% of the [OII]-detected cluster galaxies are within this lower mass range, and therefore primarily follow this correlation. In order to visually highlight the correlations present, we apply a two-dimensional kernel density estimator to all the data points, and to the lower mass galaxies.  We smooth the data using a Gaussian kernel with the width given by Silverman's rule, proportional to the standard deviation along both axes.  The FWHM of the Gaussian kernel in $\log[(r/r_{200})\times(\Delta v/\sigma_v)]$ and $D_n(4000)$ is $[0.76, 0.38]$ and $[0.60, 0.30]$ for all galaxies and lower mass galaxies, respectively.  Regions representing 68\% and 95\% of the kernel-convolved surface density are shown by the gray and and green contours, for all galaxies and the lower mass galaxies, respectively.

\subsection{Infrared Spectral Energy Distributions}
\label{sec:sed}
With broad characterization of the thermal portion of the spectral energy distribution, we can fit modified blackbodies to the stacked fluxes to estimate the dust temperature and infrared luminosity of each population.  We adopt the modified blackbody given by: 
\begin{equation}
S_{\nu}\  = A[1-e^{-\tau_\nu}]B_{\nu}(T_d)
\label{eqn:modbb}
\end{equation}
where $A$ represents an amplitude parameter, $B_{\nu}(T_d)$ is the Planck function, $\tau_\nu = (\nu/\nu_0)^{\beta}$, and we assume a crossover frequency of $\nu_0=3$ THz \citep{Blain03}.
 We keep the slope of the emissivity, $\beta$, fixed to a typical value of 1.7---with the accepted range of values between 1.5--2.0 \citep{Dunne00, Boselli12}.  We note that fixing $\beta$ at either limit does not alter the resulting dust temperatures beyond their 1$\sigma$ uncertainties.  There is also a strong degeneracy between dust temperature and redshift, as both parameters alter the location of the thermal peak in a similar manner: colder temperatures and increasing redshift shift the peak to longer wavelengths and attenuated fluxes.  However, stacking on spectroscopically confirmed star-forming members removes this problem as we no longer need to fit for the redshift, reducing the final errors on the dust temperature. 

Rather than using a predefined fitting routine, we explore a wide range of parameter values for both the amplitude and dust temperature.  We create a grid of linearly-spaced dust temperatures from 10--100\,K and logarithmically-spaced amplitudes from 0.01--1000.  For each pair of parameters ($A, T_d$), we compute the $\chi^2$ value between the model corresponding to these parameters and our stacked fluxes, summing over the five wavelengths.  

\begin{figure*}[] \centering
\includegraphics[width=13cm]{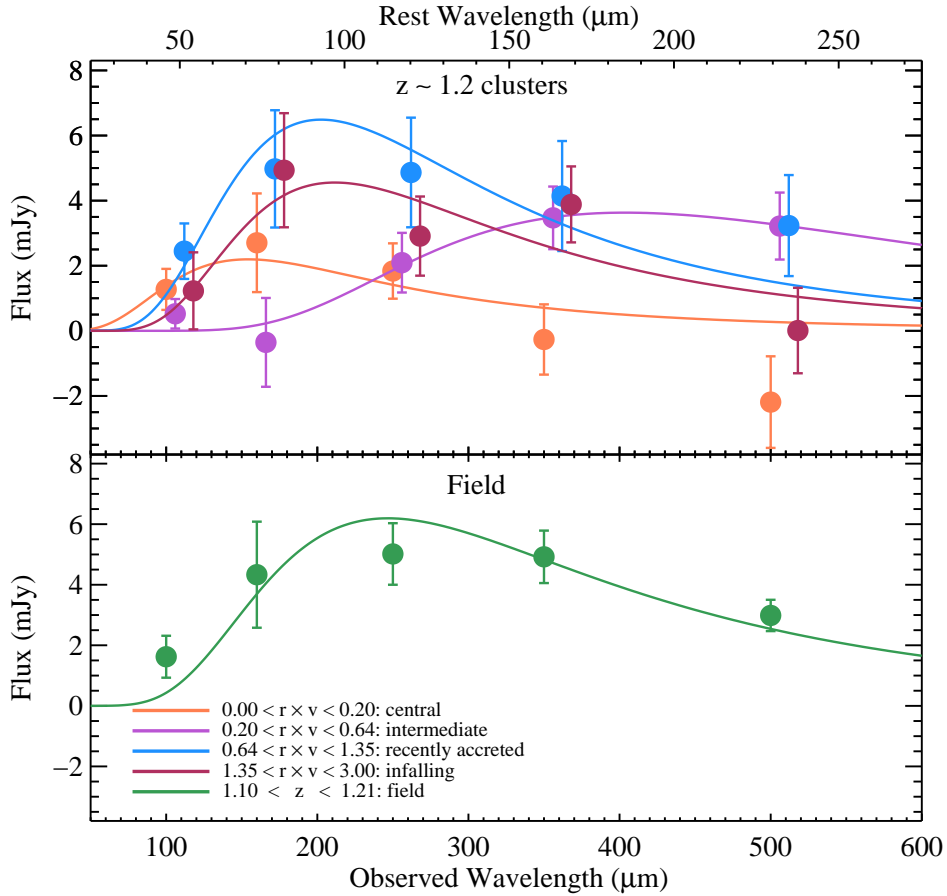}
\caption{Modified blackbody fits to the thermal portion of the spectral energy distribution using five wavelengths from stacked \textit{Herschel}-PACS (100 and 160\,\um) and \textit{Herschel}-SPIRE (250, 350, and 500\,\um) fluxes, shown as solid circles.  The best fit for each phase-space bin is represented by the solid line.  The orange, purple, blue, and maroon colors correspond to the ascending phase-space bins: $(r/r_{200})\times(\Delta v/\sigma_v) <0.20$; $0.20<(r/r_{200})\times(\Delta v/\sigma_v) <0.64$; $0.64<(r/r_{200})\times(\Delta v/\sigma_v) <1.35$; and $(r/r_{200})\times(\Delta v/\sigma_v) >1.35$, respectively.  The bottom panel is the same, except for star-forming field galaxies over $1.10<z<1.21$ (green curve). Uncertainties are measured from the standard deviation of 1000 bootstrapped stacked fluxes.  The points have been offset slightly in observed wavelength to avoid overcrowding.  There is a clear difference in the shape of the SED for the intermediate (purple) bin, which peaks at colder dust temperatures.}
\label{fig:sed_fits} 
\end{figure*}

In Figure~\ref{fig:sed_fits}, we plot the stacked fluxes and resulting best-fit modified blackbody from $\chi^2$-minimization for each of the phase-space bins (top panel) and star-forming field galaxies (bottom panel).  The intermediate phase-space bin peaks at longer wavelengths (i.e., colder dust temperatures) compared to the other bins. 

\begin{figure*}[] \centering
\includegraphics[width=17.2cm]{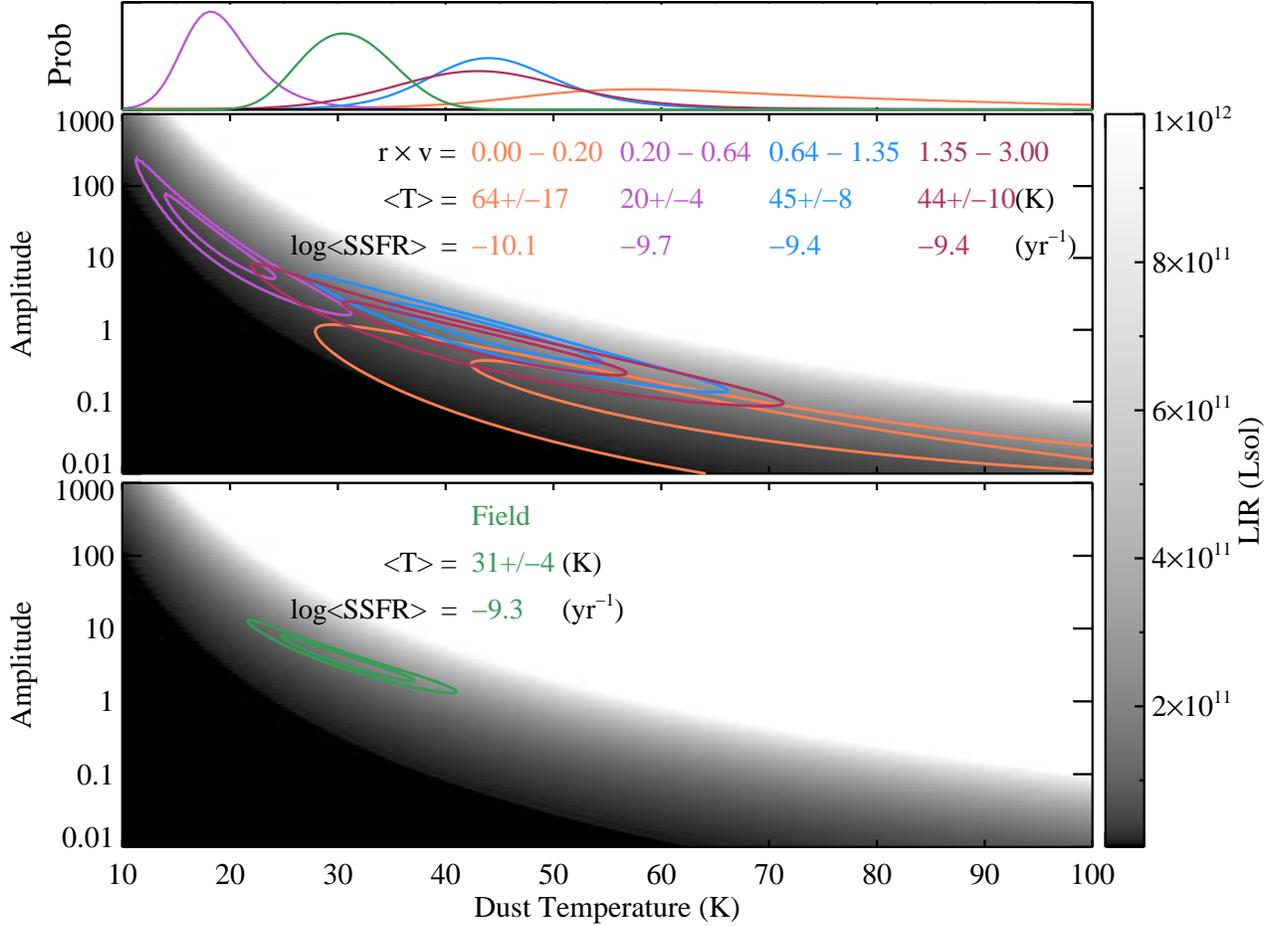}
\caption{The full parameter space of modified blackbody amplitude and dust temperature for the four phase-space bins (middle panel) and field sample (lower panel).  The gray-scale background corresponds to values of infrared luminosity, given each amplitude and temperature.  The contours represent the 68\% and 95\% likelihoods for each phase-space bin.  The upper panel shows the 1D probabilities of dust temperature, marginalized over all amplitudes.  The listed mean dust temperatures and SSFRs (derived from $\langle L_{\rm{IR}}\rangle$) are calculated using the normalized probability distribution as a weight.  The resulting error bars are given by the rms.  The intermediate phase-space (purple) bin occupies a completely separate region in parameter space compared to the earliest accreted (orange) bin.}
\label{fig:chi_prob} 

\end{figure*}

From the 2D grid of $\chi^2$ values, we further define a probability distribution, given by $P(A, T_d) \propto \exp(-\chi^2(A, T_d)/2)$.  This allows for a more accurate representation of the uncertainties given the shape of the degeneracy between $A$ and $T_d$.  Figure~\ref{fig:chi_prob} displays the 2D parameter space with contours which enclose the 68\% and 95\% surfaces of these probabilities.  We then calculate mean dust temperatures for each of the four phase-space bins (middle panel) and the field sample (lower panel), weighting by the probability distributions.  The reported values are given by, $\langle T_{d}\rangle = \int dA\, dT_d\ P(A, T_d)\, T_d$.  We similarly derive errors using the variance, $\sigma = \sqrt{\langle T_{d}^2\rangle - \langle T_{d}\rangle^2}$.  The 1D distributions for dust temperature, after marginalizing over the amplitude, are shown in the top panel of Figure~\ref{fig:chi_prob}.  

We also attempt to calculate the best fit SED for a bin of all cluster members without [OII]-emission.  The stacked fluxes in this bin are all consistent with 0\,mJy at the 1$\sigma$ level.  The concern is that if we see emission in this bin, it could be due to increased confusion in the cluster center from un-detected sources as the cluster source density rises.  This could artificially boost emission in the other bins as well, especially the central phase-space bin where the source density is highest.  This problem is somewhat mitigated by stacking all cluster members and a catalog of 24\um\ prior positions simultaneously, but only accounts for galaxies detected above $3\sigma$, not a population below the noise (at 24\um).  We are unable to constrain the 68\% and 95\% contours in the 2D parameter space for this bin (i.e., the modified blackbody template is not a good fit), and therefore conclude that higher confusion in the cluster is not significantly altering our results.

The infrared luminosities are obtained by integrating the modified blackbody over rest-frame 8--1000\,\um, and then weighted by the probability distribution. Infrared luminosities derived from fits to spectral energy distributions that span the peak of the thermal emission should provide robust estimates of the bolometric SFR within galaxies \citep{Kennicutt09}, as they trace dust emission both directly and indirectly related to star formation.  The former tracer includes any dust that is directly heated from ultraviolet radiation from young stars, while the latter describes the inherent association between dust and gas---the raw fuel for star formation. The SFRs are calculated using the relation from \cite{Kennicutt98}, and adjusted to Chabrier IMF with a factor of 1.65 \citep{Raue12}.

While the two highest phase-space bins and the field bin are entirely consistent within 1$\sigma$, there is a clear break in the full parameter space between the lowest phase-space bin (orange curve) and the intermediate bin (purple curve), at a level $>2\sigma$.  The derived properties and fit parameters are listed in Table~\ref{tab:sed_fit}.

\begin{table*} 
\begin{center}
\caption{Weighted mean and best fit values for stacked star-forming cluster members in phase-space bins.}
\label{tab:sed_fit}
\begin{tabular}{ccccccccc}
\hline
\multicolumn{1}{c}{Phase Space} &
\multicolumn{1}{c}{Curve} &
\multicolumn{1}{c}{Average} &
\multicolumn{1}{c}{$\chi^2/\nu$} &
\multicolumn{1}{c}{$\rm T_{\rm best\ fit}$} &
\multicolumn{1}{c}{$\langle\rm T_{\rm dust}\rangle$} &
\multicolumn{1}{c}{$\langle\rm L_{\rm IR}\rangle$} &
\multicolumn{1}{c}{$\langle\rm SFR\rangle$} &
\multicolumn{1}{c}{$\langle\rm log(SSFR)\rangle$} \\
\multicolumn{1}{c}{Bin} &
\multicolumn{1}{c}{Color} &
\multicolumn{1}{c}{$z$} &
\multicolumn{1}{c}{} &
\multicolumn{1}{c}{(K)} &
\multicolumn{1}{c}{(K)} &
\multicolumn{1}{c}{($10^{10}$\Lsol)} &
\multicolumn{1}{c}{(\myr)} &
\multicolumn{1}{c}{(yr$^{-1}$)} \\

\hline

\vspace{2mm}

$0.00<r\times v<0.20$ & orange & 1.179 & 2.1 & 58 & 64$\pm$17 & 9.6$\pm$4.3 & 10$\pm$4.5 & $-10.1^{+0.16}_{-0.26}$ \\ \vspace{2mm}
$0.20<r\times v<0.64$& purple & 1.179 & 0.8 & 18 & 20$\pm$4.0 & 6.6$\pm$1.5 & 6.9$\pm$1.6 & $-9.68^{+0.09}_{-0.11}$ \\ \vspace{2mm}
$0.64<r\times v<1.35$& blue & 1.178 & 1.2 & 44 & 45$\pm$7.5 & 24$\pm$5.0 & 25$\pm$5.2 & $-9.40^{+0.08}_{-0.10}$ \\ \vspace{2mm}
$1.35<r\times v<3.00$ & maroon & 1.178 & 1.6 & 43 & 44$\pm$9.6 & 17$\pm$5.3 & 17$\pm$5.5 & $-9.30^{+0.12}_{-0.17}$ \\
\hline
field & green & 1.133 & 2.6 & 31 & 31$\pm$4.0 & 17$\pm$3.6 & 17$\pm$3.8 &  $-9.34^{+0.09}_{-0.11}$ \\

\hline
\end{tabular}
\end{center}
\end{table*}

\subsection{Running Average}

\begin{figure}[h!] \centering
\includegraphics[width=8.5cm]{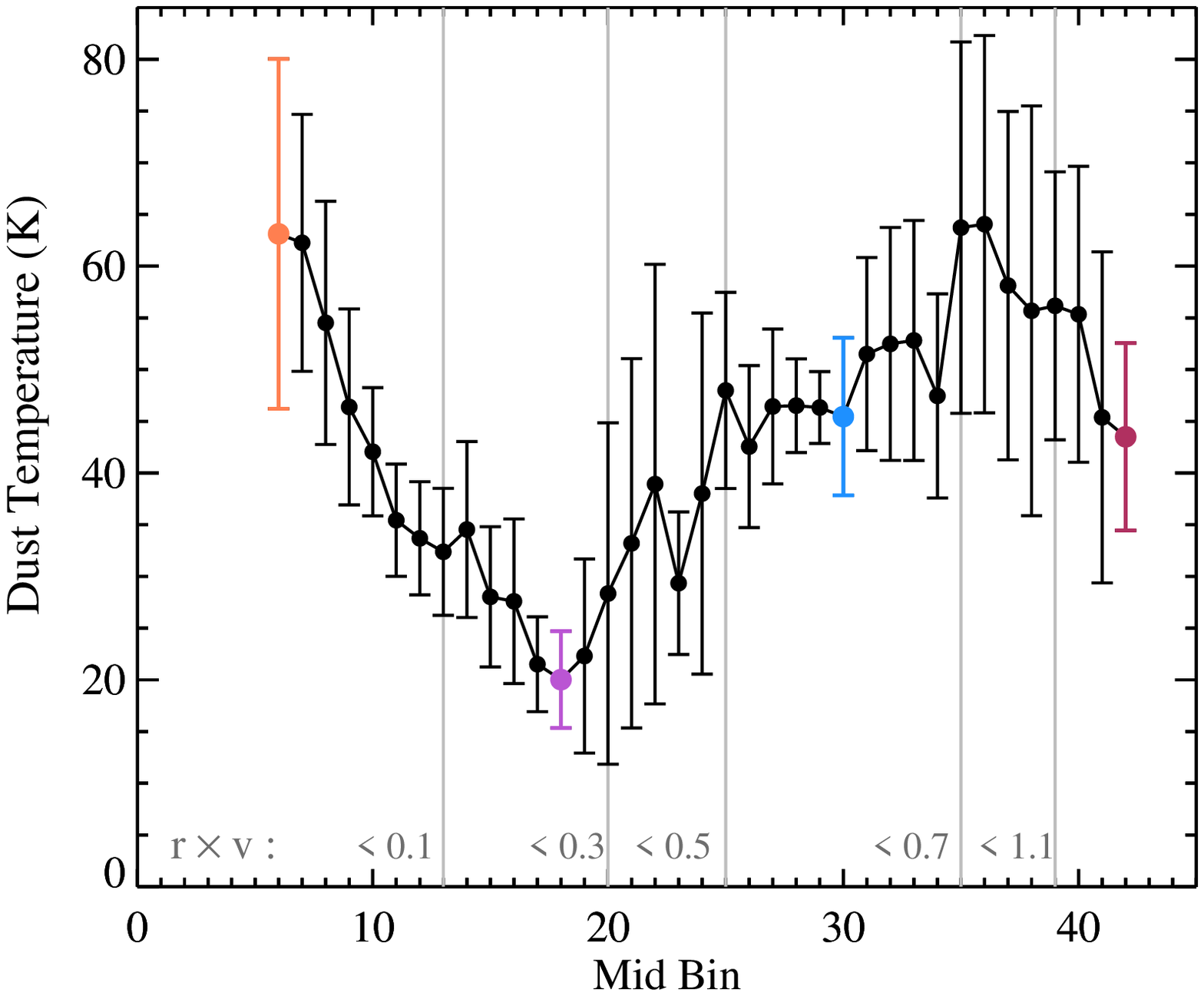}
\caption{The weighted mean dust temperature as a function of the mid bin from the running average.  The galaxies are sorted by ascending values of \rxv\, and each bin is shifted by one galaxy, for 12 binned galaxies in total.  The four independent bins from our analysis are highlighted with their appropriate colors from Figure~\ref{fig:phase}.  Arbitrary values of \rxv\ are plotted as vertical gray lines, showing the last bin that contains galaxies below that value of \rxv.  The lack of a sharp transition between any of the bins signifies that there are no single galaxies responsible for the trend. }
\label{fig:running} 
\end{figure}

Given the nature of stacking, it is possible that one galaxy could significantly alter the stacked properties and lead to the dip in the intermediate bin and/or high dust temperatures in the earliest accreted bin.  We test this by performing a running bin average, where each subsequent bin (sorted by phase-space values) replaces the lowest \rxv\ galaxy with the next galaxy in the list, yielding 36 (non-independent) bins in total.  The full probability analysis is executed for each running bin, with the weighted mean dust temperature plotted in Figure~\ref{fig:running}.  There is a smooth decline in dust temperature from the lowest phase-space bin toward the intermediate bin.  This suggests that the drop in dust temperature is due to the overall population of galaxies as they move toward higher \rxv\ values, rather than a single outlier galaxy which would manifest as a sharp drop in the running mean.

We also note that our sample of [OII] emitters contains two of the three brightest cluster galaxies (BCGs) with the lowest values of \rxv.  This is not surprising given that beyond $z\sim1$ BCGs are more likely to contain star formation as seen at 24\um\ (Webb et al.,\ submitted).  Again, we can confirm these two BCGs are not single-handedly augmenting the dust temperature in the earliest accreted bin as there is a steady change in the running average.  In fact, the third point in Figure~\ref{fig:running} roughly corresponds to the dust temperature without the two BCGs (though technically this bin also includes two additional sources that are normally in the intermediate bin since the number of sources in each bin remains at a constant 12).  If we adopt this value to compare to the intermediate bin, the significance of the change in dust temperature between the intermediate (purple) and earliest accreted (orange) bins remains unchanged at $\sim2.7\sigma$ (see \S\ref{sec:ssfr}).

\subsection{Properties of Star-forming Cluster Members at $z\sim1.2$ as a function of Phase Space}
\label{sec:ssfr}
\begin{figure}[h!] \centering
\includegraphics[width=8.5cm]{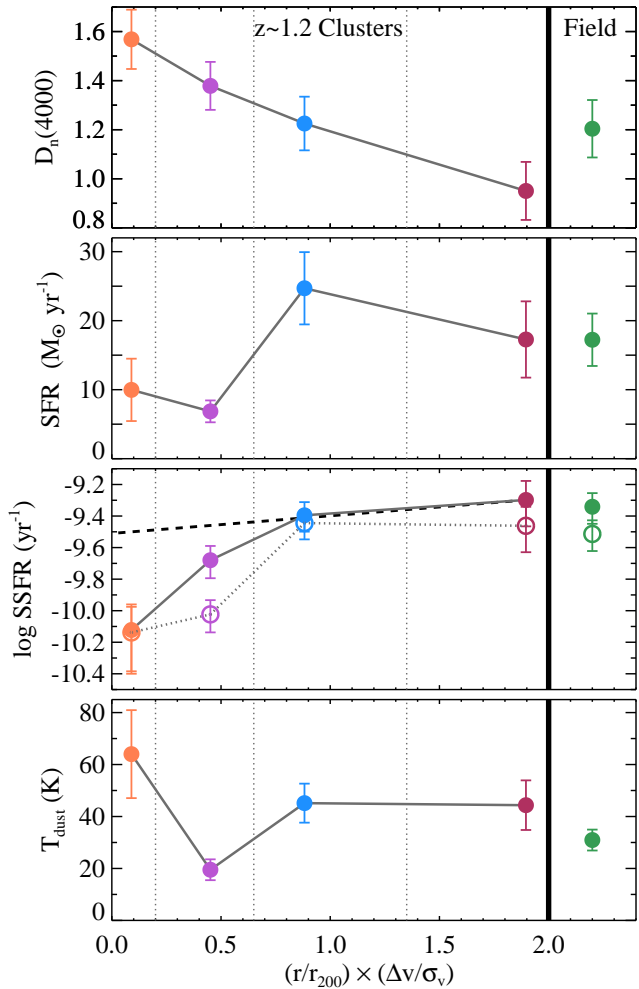}
\caption{The depth of the 4000\,\AA\ break (upper panel), SFR and SSFR (middle panels), and dust temperature (lower panel) for star-forming galaxies as a function of phase space (solid points).  The corresponding field values are shown in the right panel (green circles).  The 4000\,\AA\ break is measured on stacked spectra, with uncertainties estimated from 1000 bootstrapped resamplings in each bin.  The lower three panels are weighted means measured from the full probability distribution (see Figure~\ref{fig:chi_prob}), with rms uncertainties.  The open circles represent the SFR divided by the total stellar mass from [OII] and non-[OII] cluster members.  The dashed black line in the SSFR panel corresponds to the change expected in SSFR over the stellar mass range given the $z=1$ field trend from \cite{Elbaz07}.}
\label{fig:trends} 
\end{figure}

Our main results are highlighted together in Figure~\ref{fig:trends}, which displays various star-forming galaxy properties as a function of cluster phase space (i.e., accretion history), and out to the field.  In the upper panel, we again plot the strength of the 4000\,\AA\ break versus \rxv, but only for star-forming cluster members this time.  The break is measured on the stacked spectra in each phase-space bin, with uncertainties from 1000 bootstrap resamplings in each bin.  There is a monotonic decrease in the strength of the break toward cluster members that are infalling and/or most recently accreted (blue and maroon points), both of which are also consistent with the field value (green circle).

In the two middle panels of Figure~\ref{fig:trends}, we plot the infrared luminosity-derived SFRs and SSFRs as a function of phase-space bin.  There exists two discrete levels of star formation:  $\sim20\,$\myr\ for the field and recently accreted populations, and a drop to $\sim10\,$\myr\ for the intermediate and earliest accreted populations.  After dividing by the average stellar mass in each bin, we find a decline in SSFR with accretion history, with a 0.8 dex drop between infalling galaxies and those accreted at earlier times.  This confirms our work in \cite{Noble13_240} with MIPS galaxies in a $z=0.872$ cluster, but now with a larger sample of star-forming galaxies, more robust star-formation rates, and at a higher redshift of $z=1.2$.  In open circles, we also plot the average star formation divided by the total stellar mass of all members in each bin (with and without [OII] detections).  This is more akin to measuring the fraction of star-forming galaxies.  Again, we see a sharp transition to the intermediate phase-space bin.  Given that star-forming field galaxies from $1.1<z<1.2$ have levels of star formation consistent with the recently accreted and infalling populations, this could be suggestive of environment playing an active role in the quenching of star formation, but only after a delayed time in phase space. This delayed period could represent a period of constant SFR after the initial infall time, a long fading time after a mechanism begins to quench star formation, or a combination thereof.  However, we need to further control for stellar mass to disentangle any effects from mass-quenching in order to elucidate the precise role of environment.  We do note, however, that in the field at $z\sim1$ there is only a $\sim0.2$ dex drop in the SSFR of main-sequence star-forming galaxies as a function of stellar mass over our entire mass range.  We plot this as a dashed black line, calculating the SSFR for star-forming galaxies with the highest and lowest stellar masses in our sample, using the $z=1$ field relation from \cite{Elbaz07}.  To allow for the greatest degree of mass segregation in the cluster, we plot the SSFR corresponding to the lowest (highest) stellar mass at the largest (smallest) values of \rxv, maroon and orange bins, respectively.  We normalize the SSFR at low stellar masses to equal that of the maroon bin to illustrate the expected drop in SSFR as a function of \rxv\, assuming the largest possible range of stellar mass (2.1 dex).  In reality, the median stellar mass only varies by 0.5 dex from high to low values of \rxv, and this effect is even smaller.  It therefore seems unlikely that mass segregation could fully account for the 0.8 dex drop we see as a function of accretion history, but probably exaggerates the trend.

The bottom panel shows the weighted mean dust temperature for the dynamically distinct galaxy populations.  There is a $3.5\sigma$ drop in dust temperature in the intermediate phase-space (purple) bin (20\,K) compared to the average of the recently accreted (blue and maroon) bins, and a subsequent $2.6\sigma$ rise to 64\,K to the earliest accreted galaxies (orange).  The field population is consistent to within 1.9$\sigma$ of the two infalling galaxy bins.  If we make the simple assumption that there is a flat trend in dust temperature as a function of phase space and exclude the intermediate bin, we find the best fit amplitude to the remaining three cluster bins of $47\pm5.6$\,K.  This yields a $4.0\sigma$ deviation between the intermediate phase-space bin and an otherwise flat temperature distribution.

\section{Discussion}
\label{sec:disc}
Combining the best fit SEDs in Figure~\ref{fig:sed_fits}, the running mean from Figure~\ref{fig:running}, and the phase-space trends in Figure~\ref{fig:trends}, the most simplistic view of the data, moving from infalling to virialized (central) regions, is a removal of the warm dust component ($\lambda\lesssim100$\um\ rest-frame), and a subsequent reheating of the cold dust ($\lambda\gtrsim100$\um\ rest-frame).  We attempt to interpret this trend through the interplay between gas and dust within galaxies.

\subsection{Multi-component Dust and Gas Phases in the Infrared}
\label{sec:ssfr}
The gamut of dust grain sizes gives rise to many features in the infrared regime, and each dust component is sensitive to a particular heating mechanism.  The full infrared SED is thus a composite of modified blackbodies each described by a dust temperature.  Many studies have attempted to disentangle the dust temperature components with specific gas phases in the interstellar medium \citep[e.g.,][]{deJong84, Helou86, Cox86, Helou86, Bicay87, Dunne01, Boselli06, Bendo10, Bendo12, Boselli12, Galametz12, Bourne13, Bendo15}.

In the rest-wavelength regime of our sample, the dust is primarily composed of three temperature phases. Warm dust ($T\sim40$\,K) peaking at $\sim40$--100\,\um\ is associated with the younger stellar population, and therefore probes ionized gas around star-forming regions.  The atomic hydrogen in the galactic disk comprises cool dust of 20--30\,K that emits at 100\,\um.  The coldest dust ($T\sim15$\,K) traces quiescent molecular clouds, peaking at $\sim200$\um.  It is now becoming evident that the evolved stellar population also plays a role in heating the large grains associated with the cooler dust at $\lambda>160$\,\um\  \citep[e.g.,][]{Bendo12}.  Nevertheless, many recent studies have found that far-infrared wavelengths also correlate with direct tracers of star formation and gas, indicating that the cold dust is at least partially heated from younger stars, or probes the density of gas that fuels star formation \citep[e.g.,][]{Verley10, Galametz10, Boquien11, Bourne13, Kirkpatrick14}. 

We note that, ideally, a multi-temperature modified blackbody is required to fully understand the contributions from each component.  However, due to lack of available photometry at high-redshift, we risk overfitting the data and therefore must rely on an average dust temperature.  Moreover, there is a degeneracy between estimates of warm and cold dust temperatures from a two-temperature model \citep{Kirkpatrick12}, which could further confuse any interpretation.  This therefore warrants caution in comparing our absolute dust temperatures and SFRs to other studies with multiple dust components, and instead focus on the relative differences within our own sample.  Nevertheless, our measured dust temperatures are consistent with other studies that adopt a single-temperature modified blackbody.  For example, \cite{Hwang10} and \cite{Elbaz11} measure temperatures of $15\lesssim T\lesssim60$\,K for field galaxies out to $z\sim2.8$, with galaxies in lower-redshift clusters ($z<0.25$) displaying a similar range of values from $10<T<70$\,K though mostly concentrated between $20-45$\,K \citep{Pereira10, Auld13, Davies14}.

\subsection{A Simple Interpretation for Quenching}
We attempt to explain the above star formation and dust temperature trends with a simple interpretation of the possible quenching mechanisms at work, in light of the various dust phases.  Starting from the cluster infall regions moving to the intermediate (purple) phase-space bin, we see a $\sim3.5\sigma$ drop in dust temperature and a sharp decline in the star-formation rate.  The SED of the intermediate (purple) bin in Figure~\ref{fig:sed_fits} displays a striking lack of warm and cool dust components at rest-frame wavelengths shorter than 100\um, with only the coldest dust (represented by the SPIRE emission) dominating.  In general, some cold dust is heated by the interstellar radiation field, thereby complicating SFR estimates from luminosity-weighted blackbodies.  However, there is recent evidence that a substantial fraction of cold dust is also heated by ongoing star formation, given a strong trend between the cold dust temperature and SFR normalized by the dust mass surface density \citep{Clemens13, Kirkpatrick14}.

Progressing to the central (orange) phase-space bin, we find another change in the mean dust temperature, climbing to 64\,K from the intermediate bin at a level of 2.6$\sigma$ (but only a 1.1$\sigma$ change from the two infalling bins).  The central (orange) SED in Figure~\ref{fig:sed_fits} is devoid of the coldest dust emission at longer wavelengths, and displays scaled-down warm dust emission at rest-frame $\lambda<100$\um.  While the SFR remains roughly constant in these two interior bins (orange and purple), the SSFR drops an additional 0.4 dex.

Many mechanisms have been invoked to explain the evolution of cluster galaxies as they fall into the cluster potential, such as strangulation \citep{Larson80, Balogh00}, ram-pressure stripping \citep{Gunn72, Quilis00}, galaxy harassment \citep{Moore96}, and viscous stripping \citep{Nulsen82}. Ram-pressure stripping is a more violent process, operating deep within the cluster potential \citep{Treu03} and occurring when high-velocity galaxies move through the dense intracluster medium.  The intracluster medium (ICM) exerts a strong dynamical pressure on these galaxies, capable of stripping the cold disk of gas, thereby directly exhausting the fuel for star formation.  Molecular clouds, however, are impenetrable to the effects of stripping due to their higher densities.  Ram-pressure stripping is thus thought to have a suspended, then rapid ($\lesssim100$\,Myr), effect on star formation \citep{Wetzel12, Wetzel13}; molecular clouds already in existence can still form stars, but the removal of atomic disk gas prevents further production of molecular hydrogen. Star formation thus ceases on the scale of a molecular cloud lifetime (approximately tens of millions of years; \citealp{Blitz80}).

The phase-space trends in Figure~\ref{fig:trends} seem consistent with a delayed timescale for cluster-specific processes (or a long combination of delay plus fading time), as both the SFR and dust temperature remain constant until \rxv\ $ <1.0$.  At this point, many of the galaxies would have likely encountered the denser regions of the ICM, either on their first infall or on their way back out.  We note that statistically, the intermediate (purple) phase-space bin also contains a higher percentage of backsplash galaxies  \citep{Balogh00} than the other bins using the regions of \cite{Mamon04}.  Many of the galaxies in this intermediate (purple) bin (either those approaching pericenter or a backsplash population) are thus susceptible to ram-pressure stripping, which can effectively remove the entire HI component in the disk, along with the hot halo gas.  The dust could be stripped concurrently as spiral galaxies are likely to contain diffuse ionized gas in their halo that is associated with the presence of dust \citep{Howk99, Menard10}.  Moreover, observational evidence for dust stripping in galaxies with likely ongoing ram-pressure stripping has been detected \citep{Crowl05, Cortese10_ngc, Sivanandam14}. Therefore, the warmer dust component could be removed with the gas, along with the cool dust associated with the HI  \citep{Boselli06,Cortese10}.  As the molecular clouds are left unperturbed, the remaining dust would be the coldest component associated with dense quiescent molecular clouds.  This is corroborated in the intermediate (purple) SED, where only the coldest dust emission remains, and there is a rapid drop to low dust temperatures in this phase-space region.  Over time, these surviving molecular clouds could eventually fragment and form additional stars.  This is consistent with the residual star formation in the central (orange) phase-space bin, and the reheating of the cold dust as the new stars ionize the gas and heat the surrounding dust.  Our results thus seem consistent with a scenario in which the cluster has little effect on star-forming galaxies until they reach a sphere of influence, possibly within higher ICM densities where it is conducive to ram-pressure stripping.  At this point, there is a rapid drop in dust temperature and SFR.

Granted, we are intentionally selecting star-forming galaxies as determined by [OII] emission.  We thus should expect to find some level of dusty star-formation in each phase-space bin.  The more interesting observation is that we see evidence for a change in the level of star formation activity and dust temperature as a function of time-averaged density (i.e., phase space). Moreover, this seems consistent with ram-pressure stripping only occurring at later accretion times in the intermediate (purple) phase-space bin.  If many of these galaxies are backsplash or close to pericenter, it suggests that roughly one cluster crossing is required before any quenching effects are observed.  This corresponds to $\sim1$\,Gyr in our cluster sample.

\subsection{Dust Properties and Quenching of Cluster Galaxies in the Literature}
With the advent of \textit{Herschel}, many studies have now begun to investigate the dust properties of galaxies as a function of environment, for example, the Herschel Virgo Cluster Survey \citep{Davies10}, the Herschel Reference Survey \citep{Boselli10}, the Local Cluster Substructure Survey \citep{Smith10} and the Herschel-Astrophysical Terahertz Large Area Survey \citep{Eales10HATLAS, Dunne11}, albeit all at lower redshifts than presented here.  Almost all these studies have found a morphological \citep{Smith12, Auld13} and/or environmental dependence \citep{Smith10_Loc, Cortese12, Pappalardo12, Agius15} on the properties of dust.  Specifically, extra-planar dust existing off the disk has been found to be coincident with stripped gas in cluster galaxies  \citep{Gomez10, Cortese10_ngc}, and the extent of dust in the disk is truncated in HI-deficient Virgo galaxies \citep{Cortese10}.  These results all seem to suggest that environmental mechanisms are capable of altering the distribution of dust and/or stripping it along with the gas; in most cases, ram-pressure stripping is invoked to explain the trends.  Moreover, while \cite{Cortese12} find dust stripping from the disk to occur in star-forming cluster galaxies, they find it to be less efficient than the removal of atomic gas.  This can be explained with the presence of a less-extended dust disk that instead follows the more compact molecular gas phase, and is consistent with our findings of only the coldest dust remaining in the intermediate (purple) phase-space bin as a result of ram-pressure stripping.

Our simple interpretation of ram-pressure stripping as a viable quenching mechanism is compatible with other studies.  \cite{Wetzel12} find a bimodal distribution of SSFRs, split between central and satellite galaxies in $z\simeq0$ group and cluster galaxies from the SDSS.  They claim that environment has little effect on satellite galaxies until they cross the virial radius.  They derive quenching timescales to explain this trend \citep{Wetzel13}, finding a plausible scenario in which satellite galaxies undergo delayed ($\sim$2--4\,Gyr), then rapid ($<0.8$\,Gyr) quenching, likely due to ram-pressure stripping.  Similarly, Balogh et al.\ (submitted) propose both a long delay timescale and shorter fading time that are dependent on stellar mass, though suggest that overconsumption might be the primary cause for quenching of high SFRs in groups and clusters at $z=0.9$.  In a large sample of clusters over $0.3<z<1.5$, \cite{Alberts14} propose multiple mechanisms for altering SFRs, with  strangulation working in higher-mass galaxies and ram-pressure stripping causing a quick transition in the fraction of star-forming galaxies within the clusters at all redshifts. Moreover, direct evidence for ram-pressure stripping---in the form of trailing HI gas tails---has previously been observed in the Virgo cluster \citep{Chung07} and in a merging cluster at $z=0.3$ \citep{Owers12}.
 
Further support for mechanisms which quench star formation rapidly after a delayed period emerges with the observed abundance of poststarburst galaxies in clusters compared to the field \citep[e.g.,][]{Dressler99, Poggianti99, Poggianti09, Tran04, Tran07, deLucia09, Muzzin12}.  These galaxies exhibit strong Balmer absorption (e.g., H$\delta$) and weak emission lines (e.g., little-to-no [OII] emission), indicative of star formation that ended abruptly within the last few hundred million years \citep{Dressler83, Couch87}.  A population of poststarburst galaxies in cluster cores is expected from ram-pressure stripping given the resulting removal of disk gas, a possible compression of molecular clouds from the shock, and a subsequent rapid decline in star formation on the order of molecular cloud lifetimes.  

\cite{Muzzin14} have recently uncovered a correlation in the location of poststarburst galaxies with phase-space in SpARCS/GCLASS clusters, finding them preferentially located in an intermediate phase-space bin with higher line-of-sight velocities.  Their star formation history is best reproduced with a model that adapts a long delay period ($\sim2$\,Gyr) before a short (0.4\,Gyr) quenching of star formation.  \cite{Jaffe15} also find a segregation in phase space for HI-detected galaxies in a $z=0.2$ cluster.  They infer the presence of a stripping region in phase-space at low radii and/or high-velocities where ram-pressure can sufficiently remove the HI gas and lead to a decay in SFRs.  Our \textit{Herschel} study corroborates these findings as we measure a trend in dust temperature that is consistent with a complete removal of all warm and cool (ISM) dust after first infall that can be explained with ram-pressure stripping.

In the work presented here, we have specifically selected galaxies with [OII] emission in order to expose environmental trends with star-forming galaxies; therefore, we are not likely observing a current poststarburst population.  However, we could be witnessing the central and/or intermediate-bin galaxies during the delayed period sometime after ram-pressure stripping.  While they do exhibit somewhat depressed levels of star-formation compared to their recently accreted counterparts, they certainly are not a quenched population.  Perhaps these are the progenitors of poststarburst (i.e., recently quenched) galaxies before a rapid suppression of star formation.  Some authors have even suggested that poststarburst galaxies could be the descendants of e(a) galaxies: dusty starbursts with both moderate emission lines and strong Balmer absorption \citep[e.g.,][]{Poggianti99, Balogh05, deLucia09}.  These galaxies are thought to contain a multi-phase dust distribution that obscures young OB stars in HII regions while leaving A stars relatively unaffected \citep{Poggianti00}.   We would thus be observing these galaxies in the calm before the storm---a delayed period of somewhat passive evolution, while there is still ample molecular gas available to fuel star formation before a more rapid quenching commences.

\section{Conclusions}
\label{sec:conclusions}
We have presented a \textit{Herschel} study of star-forming galaxies in three $z\sim1.2$ clusters from the SpARCS-GCLASS sample.  We stack PACS (100 and 160\,\um) and SPIRE (250, 350, and 500\,\um) maps at the location of galaxies with [OII] emission---the star-forming population---binned by their location in $(r/r_{200})\times(\Delta v/\sigma_v)$ phase space based on our previous study in \cite{Noble13_240}.  We utilize bins with an equal number of 12 star-forming galaxies: $(r/r_{200})\times(\Delta v/\sigma_v) <0.20$ (central bin); $0.20<(r/r_{200})\times(\Delta v/\sigma_v) <0.64$ (intermediate bin); $0.64<(r/r_{200})\times(\Delta v/\sigma_v) <1.35$ (recently accreted bin); and $(r/r_{200})\times(\Delta v/\sigma_v) >1.35$ (infalling bin). This isolates the earliest accreted cluster galaxies from galaxies that have completed at least one passage through the cluster, and those that are currently infalling.  We summarize our results as follows:
\begin{enumerate}

\item{We fit the thermal portion of the stacked spectral energy distribution for each phase-space bin, deriving weighted mean dust temperatures and integrated infrared luminosities using photometry from 100--500\,\um\ in the observed frame, corresponding to 45--230\,\um\ rest-frame.  After converting the luminosities to SSFRs, we find a steady 0.8 dex decline from the infalling population towards virialized (central) galaxies in the core.  This confirms the MIPS study of a $z=0.871$ cluster presented in \cite{Noble13_240}.  The actual SFRs and the fraction of star-forming galaxies, however, display a rapid decline in the intermediate phase-space bin, suggesting a rapid quenching of cluster galaxies.}

\item{There exists a $\sim$4.0$\sigma$ drop in the dust temperature for the intermediate phase-space bin, when compared to a flat trend that is fit to the infalling and central bins.  Its full probability distribution favors cooler dust temperatures and occupies a distinct region in the 2D parameter space of temperature and amplitude for a modified blackbody. The recently accreted/infalling galaxies have both dust temperatures and SSFRs consistent with the $1.10<z<1.21$ field galaxies, suggesting they have not undergone any substantial evolution within the cluster yet.}

\item{A running bin average of weighted dust temperatures shows a removal of warm dust moving inwards in phase space, from the infalling galaxies to the intermediate phase-space bin.  There is then a steady rise in the dust temperature towards the virialized (central) galaxies.}

\item{We propose a simple interpretation for quenching in which infalling galaxies remain unscathed until roughly one cluster crossing ($\sim1$\,Gyr).  At that point, they experience the violent stripping of all dust and gas components, except in the densest regions of quiescent molecular clouds which contain the coldest dust.  As the surviving clouds form stars, there is a reheating of the coldest dust in the earliest accreted star-forming galaxies.}

\end{enumerate}

We emphasize that this last conclusion is just a plausible quenching model to explain the observed trends in star formation activity and dust temperatures as a function of phase-space environment.  We have not attempted to account for mass segregation within the clusters, which could partially contribute to the trends.  Further observations of the gas components are crucial to verify this claim; a detailed study of CO gas in cluster galaxies would provide insight into the available reservoir of molecular gas that fuels star formation.  However, qualitatively, it illustrates the power in the parameterization of \rxv\ in phase space as a means for studying galaxy evolution in the context of time-averaged densities.

\acknowledgments
We thank the anonymous referee, whose comments improved the clarity of the manuscript.  The authors would also like to thank numerous people for useful discussions, including Rachel Friesen, Suresh Sivanandam, Alexander van Engelen, and Marco Viero.  This work is based in part on observations made with \textit{Herschel}, a European Space Agency Cornerstone Mission with significant participation by NASA. Support for this work was provided by NASA through an award issued by JPL/Caltech.
TMAW acknowledges the support of the NSERC Discovery Grant.  HKCY is supported by the NSERC Discovery Grant and a Tier 1 Canada Research Chair.  GW gratefully acknowledges support from NSF grants AST-0909198 and AST-1517863.  RFJvdB acknowledges support from the European Research Council under FP7 grant number 340519.

\textit{Facilities:} \facility{Herschel Space Observatory (PACS; SPIRE)}, \facility{Spitzer Space Telescope (MIPS; IRAC)}, \facility{Gemini (GMOS)}

\bibliography{references}
\bibliographystyle{apj}

\end{document}